\documentstyle{l-aa-ps}
\input{psfig}
\begin{document}

\def\Su{$u$}
\def\Sv{$\nu$\/ }
\def\Sb{$b$}
\def\Sy{$y$}
\def\Suv{($u$--$\nu$)}
\def\Svb{($\nu$--$b$)}
\def\Sby{($b$--$y$)}
\def\Ha{H$\alpha$}
\def\Hb{H$\beta$}
\def\Hg{H$\gamma$}
\def\Gr{{\it r}\/}
\def\Gi{{\it i}\/}
\def\UB{(U--B) }
\def\BV{(B--V) }
\def\BR{(B--R) }
\def\VI{(V--I) }
\def\RI{(R--I) }
\def\JH{(J--H) }
\def\HK{(H--K) }
\def\HR{H$\alpha$--R}
\def\RH{R--H$\alpha$}
\def\ea{et al.\ }

\thesaurus{08(08.02.1; 08.09.2 NX Pup; 08.09.2 CG1; 08.16.5)}

\title{Low-mass star formation in CG1: a diffraction limited search for
pre-main sequence stars next to NX\,Pup
\thanks{based on observations obtained at the European Southern Observatory,
La Silla, Chile}
}
\subtitle{}

\author {Wolfgang Brandner \inst{1,2}, Jerome Bouvier \inst{3}, Eva K.\ Grebel \inst{4}, Eric Tessier \inst{5}, Dolf de Winter \inst{6},
and Jean-Luc Beuzit \inst{7}}

\offprints {W.\ Brandner}

\institute{
Astronomisches Institut der Universit\"at W\"urzburg, Am Hubland,
D-97074 W\"urzburg, Germany, brandner@astro.uni-wuerzburg.de
\and 
European Southern Observatory, Casilla 19001, Santiago 19, Chile,
\and
Laboratoire d'Astrophysique, Observatoire de Grenoble, Universit\'e J. Fourier, 
B.P. 53, F-38041 Grenoble Cedex 9, France,
bouvier@gag.observ-gr.fr
\and
Sternwarte der Universit\"at Bonn, Auf dem H\"ugel 71, D-53121 Bonn,
Germany, grebel@astro.uni-bonn.de
\and
Royal Greenwich Observatory, Madingley Road, Cambridge CB3 0EZ, England,
tessier@mail.ast.cam.ac.uk
\and
Astronomisch Instituut ``Anton Pannekoek'', University of          
       Amsterdam, Kruislaan 403, NL-1098 SJ Amsterdam, The Netherlands,
DOLF@astro.uva.nl
\and
Observatoire de Paris, DESPA (URA 264/CNRS), 5 place Jules Janssen, 
F-92195 Meudon, France, jlbeuzit@hplyot.obspm.circe.fr
}

\date {Received 25 October 1994; accepted 22 November 1994}

\maketitle

\markboth{Brandner et al.: Diffraction limited search for PMS stars near NX Pup}
{}

\begin{abstract}

Using adaptive optics at the ESO 3.6m telescope, we obtained diffraction
limited JHK-images of the region around the Herbig AeBe star NX Pup.
We clearly resolved the close companion (sep. $0\farcs 128$) to NX Pup
-- originally discovered by HST -- and
measured its JHK magnitudes. A third object at a separation of $7\farcs0$
from NX Pup was identified as a classical T Tauri star so that NX Pup
may in fact form a hierarchical triple system. 
We discuss the evolutionary status of these stars and derive estimates 
for their spectral types, luminosities, masses and ages.

\keywords{ 
Pre-main sequence evolution -- Globules: Individual: CG1 -- NX\,Pup
-- adaptive optics: diffraction limited imaging}

\end{abstract}

\section{Introduction}
Cometary Globules (CGs) are cool clouds with compact, dusty, opaque heads, 
and faint tails stretching away from the head (e.g.,
Harju \ea 1990). They were first described and defined as a group
by Hawarden and Brand (1976) based on Schmidt plates of the Gum Nebula
and NGC\,5367.
Many CGs are found in the Gum Nebula, a large region of
ionized gas at a distance of approximately 450pc. In this nebula the
isolated neutral globules clearly contrast with the surrounding hot
ionized media (Brand \ea 1983).

The heads of most CGs point to an apparent center
in the Gum Nebula, suggesting a common origin of the cometary
shape of these globules (Zealey \ea 1982).
Different mechanisms have been proposed to explain the origin of the
tail like structure of the CGs,
such as interactions of dense cloud cores with shock waves from a SN 
explosion (Brand \ea 1983) or ionization shock fronts from nearby OB
stars (Reipurth 1983). 
IRAS observations show an enhanced star formation rate
in the CGs of the Gum Nebula with respect to comparison fields
(Bhatt 1993). 
Star formation may have been induced
through shocks triggered by the same event that ionized the Gum Nebula
complex.
Reipurth and Pettersson (1993) identified eight late-type \Ha\ emission 
stars in association with CG4/CG6/Sa101 and CG13. They argue that
star formation has occurred in the Gum Nebula for more than $10^6$ years.
A mapping of CG1 in molecular lines has been carried out by Harju \ea 
(1990).
They find that CG1 is not in dynamical equilibrium and in CO shows a high 
(5km/s) velocity component near its head. They conclude that this is 
shocked material related to star forming processes.

Located at the edge of CG1 is the Herbig~AeBe star NX\,Pup (Irvine 1975). 
NX\,Pup is a variable star (Hoffmeister 1949, Strohmeier \ea 1964).
It shows \Ha\ in emission (Wackerling 1970) and has a strong UV excess
(de Boer 1977) and IR excess (Brand \ea 1983, Reipurth 1983). 
Its evolutionary status is controversial. Brand \ea and Reipurth
find an MK type of F0-2\,III, a total luminosity of 40-50 L$_\odot$, and an age
of 10$^6$ years, whereas (on the basis of UV observations from space)
de Boer (1977) and Tjin A Djie \ea (1984) assign
an MK type of A0-1\,III. Tjin A Djie \ea and Th\'e and Molster (1994) derive a total
luminosity between 100 and 160L$_\odot$ and an age of 8$\cdot$10$^5$ years for 
NX\,Pup.
The age of NX\,Pup agrees very well with the age of CG1 derived from 
dynamical studies (e.g., Harju \ea 1990).

Observations with the Fine Guidance Sensor system aboard the Hubble
Space Telescope reveal that NX\,Pup (HIC 35488) is a close
visual binary with a separation of $0\farcs 126$ (Bernacca \ea 1993).
Thus all previous evolutionary interpretations should be reevaluated as
the total observed luminosity in fact comes from two stars.

Using the ESO adaptive optics (AO) imaging system COME-ON+ (CO+) in 
combination with the SHARP camera we started a diffraction limited
imaging survey of intermediate-mass pre-main sequence
(PMS) stars listed in ``A new catalogue of members and candidate members
of the Herbig~Ae/Be stellar group'' (Th\'e \ea 1994) with the aim to search
for close IR companions. During the first part of this project we obtained
JHK images for the NX Pup system. The CLEANed JHK images clearly resolve
the close companion, here referred to as NX Pup B, detected in the optical
by the HST. 

We also found a third, faint PMS star at a distance of $7\farcs 0$ 
from NX\,Pup A/B, which will be referred to as NX\,Pup C in the following.

\begin{table}
\caption{\label{obs}Journal of observations}  
\begin{center}
\begin{tabular}{lrcc}
\hline
telescope/instr.& date & filter/wavelength\\
\hline 
ESO\,1m/vis. phot.     &1.--3.12.1992 & V  \\	
SAT (LTPV)/vis. phot.  &12.92--2.93 & y  \\	
ESO\,3.6m/CO+          & 1.1.1994 & J,H,K  \\
Danish\,1.5m/CCD-camera& 8.1.1994 & U,B,V,R,i,\Ha \\
ESO\,1.5m/B\&C         &26.1.1994 & 380-760nm \\
NTT/EMMI               &20.3.1994 & 600-725nm  \\
\hline
\end{tabular}
\end{center}
\end{table}

\section{Observations and results}

\subsection{Near-infrared Imaging}

JHK images of NX\,Pup with high spatial resolution were obtained in
1994 Jan 1 at the ESO~3.6m telescope. We used the COME-ON+
adaptive optics system (Beuzit and Rousset, 1994) in combination
with the SHARP II (System for High Angular Resolution Pictures, see
Hofmann et al. 1992)
camera from the Max Planck Institute for Extraterrestrial Physics (MPE). 
The SHARP camera
is equipped with a Rockwell NICMOS-3 array. The image scale was 
$0\farcs 050$/pixel, giving a field size of $13'' \times 13''$.
The total exposure time on NX Pup
was 300s in each JHK filter, and similar exposures were obtained on the nearby
sky for sky subtraction. Immediately after NX Pup, we observed 
a reference point source 10$'$ away (star no.985 in the HST Guide Star
Catalogue) which was later used for image deconvolution. The photometric
standard HR 3421 was observed under the same conditions to get absolute JHK
photometry. 

\begin{figure}
\centerline{\psfig{figure=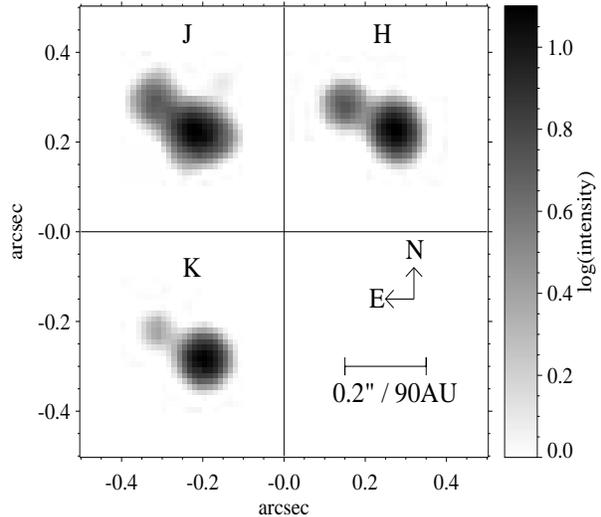,width=9.5cm,height=7.5cm}}
\caption{\label{nxco}Deconvolved JHK images of NX Pup 
obtained
with the adaptive optics system CO+ at the ESO 3.6m telescope in 1994 Jan 1.
The images were rebinned by a factor of 4 (see text for more details). 
The components A and B (sep. $0\farcs 128$, PA 62.4$^\circ$) are clearly 
resolved. At a distance of 450pc  $0\farcs 128$ correspond to a projected 
separation of 58AU. The faint feature in the J image north-west of
NX Pup A is an artefact of the image deconvolution. A
logarithmic gray scale was used. Component C 
is outside these frames.
North is up and east
is to the left.}
\end{figure}

The individual exposures in each filter, 5 in J, 150 in H, and 600 in K,
were coadded using the ``shift and add'' (SAA) technique, i.e., coadding the
individual images after shifting them so that the centre of the point spread 
functions coincide. 

\begin{table}
\caption{\label{ims}Gain in spatial resolution (FWHM and Strehl ratio, SR)
through shift and add (SAA) and SAA with 
image selection (IS).}  
\begin{center}
\begin{minipage}{70mm}
\begin{tabular}{lccc}
\hline
& J & H & K\\
\hline 
FWHM                   & $0\farcs214$ & $0\farcs172$ & $0\farcs156$ \\
FWHM (SAA)             & $0\farcs209$ & $0\farcs157$ & $0\farcs147$ \\
FWHM (SAA+IS)           & ---\footnote{no IS (only 5 individual exposures)}& $0\farcs138$ & $0\farcs143$ \\ \hline
SR                      & 4\%      & 8\%              & 25\%  \\ 
SR (SAA)                & 4\%      & 9\%              & 27\%  \\ 
SR (SAA+IS)             & ---     & 12\%              & 31\%  \\ \hline
selection rate (IS)    & ---     & 20\%              & 40\%  \\
\end{tabular}
\end{minipage}
\end{center}
\end{table}
 
All AO systems do only partially compensate the wavefront distorsions.
``The performance of an AO system is best defined in terms of the
maximum intensity obtained in a point source image. The ratio
of this intensity to the maximum intensity in the diffraction-limited
image (Airy pattern) is called the Strehl ratio'' (Roddier 1992).
The uncompensated power forms a halo around each object.

If a sufficient number ($\ge 50$) of individual exposures is available, one can gain 
in spatial resolution by selecting images 
based on a Strehl ratio (SR) criteria. Furthermore, if the single frame
exposure times are short, the residual tilt between the individual
images is reduced by applying an SAA algorithm. 
While the latter already gives some improvements mostly in the FWHM, 
the results of image selection are more significant with respect
to the Strehl ratio.
However, the rejection rate for image selection must be limited in order 
to keep a sufficient 
signal to noise ratio.
Table \ref{ims} shows the improvement in resolution through
SAA and image selection.
Out of 600 K band frames 240 were selected and coadded, improving
the Strehl ratio from 25\% to 31\%. Selecting 30 images out of the 150 H band 
exposures improved the Strehl ratio to 12\% -- compared to 8\% without 
image selection. 

All these operations were performed with the ``local'' IRAF package c128 
developed by E. Tessier at the Observatory of 
Grenoble\footnote{The IRAF package c128 is available via anonymous ftp from 
hplyot.obspm.fr. in the directory /iraf\_hra.}.
See also Tessier
et al. (1995) for a more detailed description of the data reduction .

On the shift-and-add image in the K band, NX PUP C itself appears to be 
elongated in the east-west direction 
(FWHM in east-west direction and north-south direction are
$0\farcs27 \times 0\farcs18$, compared to $0\farcs16 \times 0\farcs16$ 
for the reference source). 
This elongation is less pronounced in the H band, 
and is not seen in the J band.

NX Pup C is 7$''$ away from NX Pup AB, which was used for wave-front sensing.
Therefore, we could expect effects from
angular anisoplanatism (Wilson and Jenkins 1995).
Because of this and the lower signal to noise ratio of NX Pup C,
we did not try to deconvolve it.

In the following we will discuss briefly 
two kinds of anisoplanatism -- angular and temporal --
which cause image elongations. 
One effect of angular anisoplanatism
is that the Strehl ratio drops and the PSF gets wider with increasing
angular distance to the guide star.
The FWHM of NX Pup C is systematically wider than the FWHM of the
calibration source in each JHK filter.
This might be because of angular anisoplanatism.
Furthermore, angular anisoplanatism would lead to an elongation 
of the off-axis PSF in the direction of the guide source
(McClure et al. 1991).
In the case of NX Pup AB as reference source and NX Pup C as
programme star the direction of the elongation would be $\sim 45^\circ$,
i.e., from north-east to south-west but not in east-west direction.

Elongations in another direction can be explained by
temporal anisoplanatism due to the wind 
in the dominant turbulent layer (see Roddier
 et al. 1993 and Wilson and Jenkins 1994). 
The prevalent wind direction on La Silla near the ground is north-south,
but we do not know the wind direction 
in the dominant turbulent layer during our observations.
While an off-axis PSF should suffer from both effects,
the on-axis PSF should show pure temporal anisoplanatism.
However, the on-axis PSF of the reference source 
does not show any significant deviation from circular symmetry. 
Summarizing, neither angular nor temporal anisoplanatism are very likely 
sources of elongation and the
elongation may therefore come from the object NX Pup C itself. 
To know for sure NX Pup C would have to be re-observed. 

The resulting coadded images were then
deconvolved using a CLEAN algorithm and the nearby reference star as a PSF.
The central $0\farcs 5 \times 0\farcs 5$  part of the CLEANed JHK images is shown in
Fig. \ref{nxco}. The close visual companion detected by HST is clearly seen in
all filters.  
This is a clear demonstration of the diffraction limited imaging
capabilities of the CO+ adaptive optics system.

Astrometry has been performed on the CLEANed images, whereas photometry of 
the various components of the system has been performed
on the shift-and-add images. We did not carry out
the photometry on the CLEANed images as deconvolution algorithms
usually do not conserve the flux distribution.
 Absolute JHK photometry was obtained
for NX Pup A+B and for NX Pup C using the IRAF/APPHOT package, while
differential photometry between NX Pup A and B was performed within
the IRAF/DAOPHOT package using IRAF scripts kindly provided by A. Tokovinine.
The JHK magnitudes derived in this manner for NX Pup A, B, and
C are listed in Table \ref{nir}. The overall photometric errors are about
10\%.

\begin{table}
\caption{\label{nir} JHK photometry (1.1.1994, ESO\,3.6m/CO+). \protect\newline For comparison we also show the V magnitudes.} 
\begin{center}
\begin{minipage}{70mm}
\begin{tabular}{cccc}
\hline
filter& NX Pup A & NX Pup B  & NX Pup C \\
\hline
J       & 8\fm58 & 9\fm56 & 11\fm71 \\
H       & 7\fm43 & 8\fm37 & 10\fm66 \\
K       & 6\fm15 & 7\fm90 & 10\fm10 \\ \hline
V	& 10\fm3\footnote{assuming $\Delta$V=0\fm64 (Bernacca \ea 1993; 1.1.1993) and V=9\fm8 for NX Pup A+B} & 10\fm9$^a$ & 16\fm03\footnote{this paper (8.1.1994)} \\
\end{tabular}
\end{minipage}
\end{center}
\end{table}

\subsection{Visual Imaging}

CCD images in the Bessel UBVR, Gunn\, i, and \Ha\ filters of the region
around NX\,Pup were obtained in 1994 Jan 8 with the CCD-camera at the 
Danish 1.5m telescope at La Silla. This camera
is equipped with a 1K TEK CCD (ESO \#28) and has a scale of 
$0\farcs38$/pixel.

\begin{figure}[ht]
\centerline{\psfig{figure=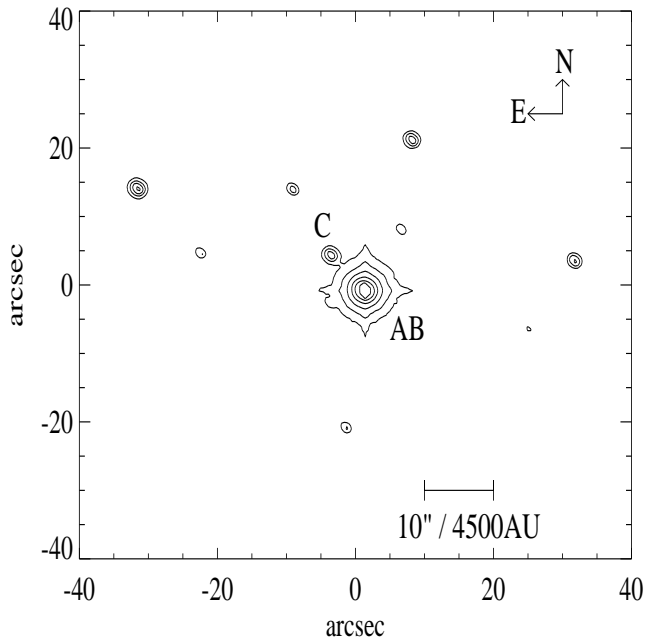,width=9.5cm,height=9.5cm}}
\caption{\label{cont}The PMS stars NX Pup A+B (center) and C in a 
10s R exposure with the CCD camera at the Danish 1.5m telescope on La Silla. 
The components AB (unresolved) and C are marked. All other stars within
the field show no excess in \Ha\ and therefore are likely to be
field stars. North is up and east is to the left.}
\end{figure}

We obtained
two sets of exposures in U,B,V,R,i,\Ha\ -- short exposures (10s, 5s, 1s, 1s, 1s, 60s) 
in order not to saturate NX\,Pup~AB
and long exposures (180s, 20s, 10s, 10s, 10s, 600s) in order to get reliable 
photometry of NX\,Pup\,C. The seeing was between $1\farcs 0$ and $1\farcs 5$.
The finding chart for NX Pup AB and C is shown in Fig. \ref{cont}.

\begin{table}[b]
\caption{\label{phot}Visual Photometry (8.1.1994, Danish\,1.54m)}
\begin{center}
\begin{minipage}{70mm}
\begin{tabular}{crr}
\hline
filter& NX\,Pup~AB & NX\,Pup~C \\
\hline
U	&10\fm40 & 17\fm52 \\
B	&10\fm30 & 17\fm46 \\
V	& 9\fm78 & 16\fm03 \\
R	& 9\fm46 & 14\fm96 \\
I	& 9\fm01 & 13\fm68 \\ \hline
\RH\footnote{see definition of \Ha \ magnitude in the text}	&0\fm46 & 0\fm68 \\
\end{tabular}
\end{minipage}
\end{center}
\end{table}

Two standard fields (SA98 and RU149) of the Landolt list 
(Landolt 1992) were observed in order to calibrate the photometric
data. 

The CCD camera at the Danish 1.5m telescope has a shutter delay
time of 30ms. For short exposures of 1sec this results in a 3\% gradient
in intensity from the center to the edge of the field.
All short exposures were corrected for the shutter delay.
Instrumental magnitudes and transformation equations were computed
and solved within the IRAF/DIGIPHOT package. The overall photometric
errors are between 2\% and 5\%. Table \ref{phot} lists the results of the
visual photometry.

As NX Pup AB is variable 
we have to be careful when combining the observations of different epochs. 
We do not have an absolute V magnitude for the HST observation of NX Pup,
but one of us (DdW) observed NX Pup with the visual photometer at the 
ESO 1m telescope between 1992 Dec 1 and 3. 
Furthermore, NX Pup is one of the target stars in the LTPV
programme and was observed between 1992 Dec 4 and 22 and
1993 Jan 21 and Feb 8.
P.S. Th\'e was so kind to
provide us Str{\o}mgren y band measurements of NX Pup -- obtained
with the Danish 50cm Str{\o}mgren Automatic
Telescope (SAT) on La Silla, Chile -- prior to their publication
(Sterken et al. 1995, in prep.). NX Pup was getting fainter
from December to February. If we assume the
y band measurements to be equivalent to the V magnitudes, simple
linear interpolation gives us V = $9\fm8 \pm 0\fm2$ at the time of the HST 
observations.
Thus NX Pup had about the same V magnitude during the CO+ observations
as during the HST observations.

\subsection{Spectroscopy}

We used the ESO\,1.5m and the Boller and
Chivens (B\&C) spectrograph in 1994 Jan 26 to get long slit spectra of
NX\,Pup\,AB and C simultaneously. The B\&C spectrograph is equipped
with a 2K
FORD chip (ESO~\#24) and can be rotated as a whole in order to align
the slit along a user defined position angle.
We used grating \#23 with an useful wavelength range from
380 to 760nm and a sampling of 0.19nm/pixel. 
The FWHM along the slit of the 2D spectra is $4\farcs3$. This relatively 
large value presumably cannot be
attributed to the seeing but to guiding errors during the 30min 
exposure. Accordingly, the shape of the one dimensional point spread function
is better described by a Gaussian than by a function resembling a King 
or Moffat profile. 
Because of the large FWHM and the large brightness
difference between NX\,Pup\,AB and C (5.5\,mag in R), the C component is
heavily blended. 

\begin{figure*}[ht]
\centerline{\psfig{figure=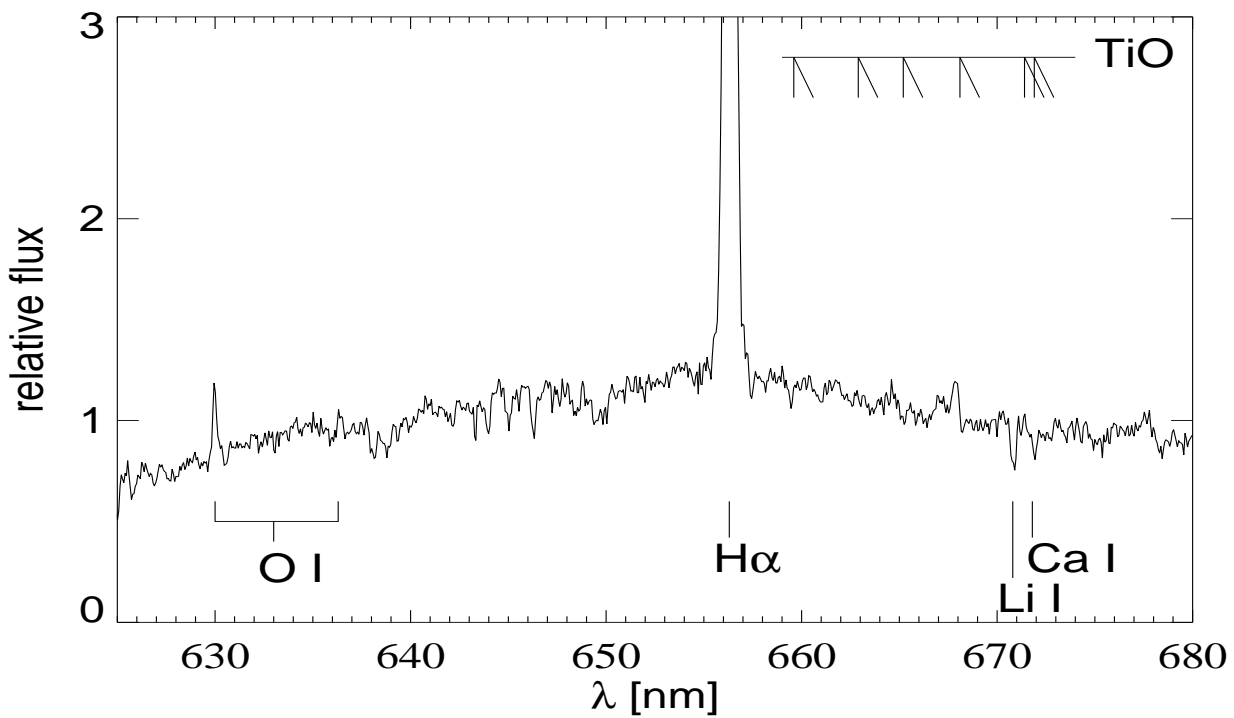,width=18.5cm,height=10.5cm}}
\caption{\label{li}Spectrum of NX Pup C, obtained in a 15min exposure with
NTT/EMMI in 1994 Mar 20. Several emission (OI, \Ha) and absorption
(TiO bands, Ca I)
features are marked. Note the strong Li I 670.8nm absorption, which is
a sign of youth.}
\end{figure*}

To extract the spectrum of NX\,Pup\,C we fit two Gaussians simultaneously
along each scan in spatial direction of our two dimensional spectra.
FWHM and separation of the two Gaussians remain fixed, allowing
only their heights and the absolute position to vary from scan to scan
in the dispersion direction. A comparison to the NTT spectrum (see
below) shows good agreement of the spectrum of NX\,Pup\,C in the overlapping
region, giving us some confidence in our method.

Spectra with a better spectral resolution (0.06nm / pixel) and spatial
resolution (FWHM $1\farcs0$)
were obtained in 1994 Mar 20 with 
the ESO Multi-Mode Instrument (EMMI) attached to the NTT. We used
the red arm of EMMI, which has a 2K TEK CCD (ESO \#36) and a pixel scale
of $0\farcs 27$/pixel.
The exposure time was 15min and the wavelength range 600 to 725nm. 
Because of the large brightness difference between NX\,Pup\,C and AB,
at least parts of the
spectra of NX\,Pup\,AB were saturated in both exposures.
The saturated regions were excluded from the fits and the final extraction.
Two spectrophotometric standard stars were observed in order to
derive the sensitivity curve and to achieve a flux calibration.
All spectra were extinction corrected applying the default La Silla
extinction curve. 

Figure \ref{li} shows the medium resolution spectrum of NX Pup C near the 
\Ha\ line.
The strong \Ha\ emission and the Li absorption can be seen.
%Line profiles of \Ha\ for NX Pup AB and NX Pup C are plotted 
%in Fig. \ref{hahb}.

\section{The evolutionary status of NX\,Pup\,A+B and C}

\subsection{NX\,Pup\,A+B}

The position angle (PA) between NX\,Pup\,AB and C measured on the CO+ frames 
and the CCD frames
is $45\fdg 3\pm 0\fdg 2$ and the separation $6\farcs 98 \pm 0\farcs 04$.
At a distance of 450 pc
this corresponds to a projected separation of 3140AU, so that NX\,Pup\,AB
and C are unlikely to form a gravitational bound system. On the other hand,
this resembles what we would expect to see for a hierarchical
triple system: two stars in a close orbit around each other and a third, less
massive star in a wide orbit. Only radial velocity measurements
and proper motion studies could tell us 
whether or not the three stars form a physical triple system.

NX\,Pup\,A and B themselves are separated by $0\farcs 128 \pm 0\farcs 008$,
as estimated from the JHK images, i.e.\ 58AU projected separation. 
They are almost certainly bound, as has been pointed out by Bernacca \ea (1993).
The separation and the PA of $62\fdg 4\pm 5\fdg 7$ are in good agreement with
the values determined by Bernacca \ea  ($0\farcs126 \pm 0\farcs007$, 
$63\fdg4 \pm 1\fdg0$).
As there is only a time span of one year between the HST measurements 
and our CO+ observations, we would not expect signs of orbital motion. 

To transform the \Ha\ magnitudes, our Bessell
R filter can be used as continuum filter. The ratio of 
our instrumental R and \Ha\ filter passbands corresponds 
to a difference of 3\fm9 (Grebel et al. 1994)
in the sense of R$_{inst}\approx $H$\alpha_{inst}-$3\fm9 (for stars without 
\Ha\ emission). A comparison of the (R$_{inst} - $H$\alpha_{inst} + 3\fm9$)
colours shows that both 
NX\,Pup~AB and C have a strong excess emission in \Ha\ (cf. Table \ref{phot}). 
No other stars with R down to 19\fm3 within a 
$2\farcm5 \times 2\farcm5$ field centered on NX\,Pup possess a significant
excess in \Ha. 

\begin{figure}[ht]
\centerline{\psfig{figure=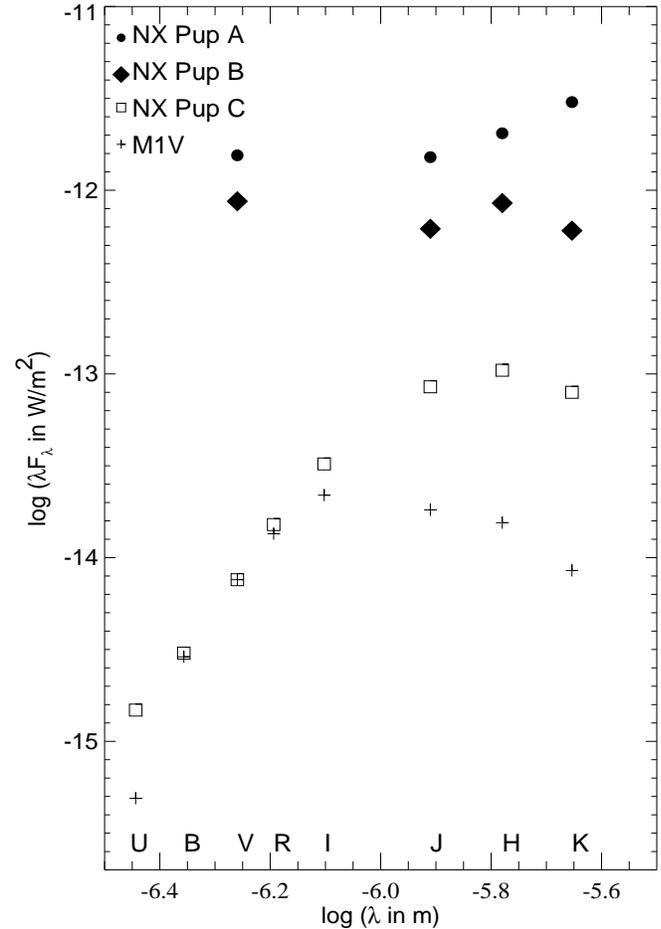,width=9.5cm,height=13cm}}
\caption{\label{sed}Spectral energy distribution $\lambda$F$_\lambda$ of
NX Pup A and B (VJHK) and
C (UBVRIJHK). For comparison we also show the spectral energy distribution
of a M1 type
star of the same apparent V magnitude as NX Pup C.
Note that the SED is rising towards longer wavelength
for NX Pup A. The SED of NX Pup B and C peaks near 1.5$\mu$m.
NX Pup C show
IR excess and UV excess.
The errors in flux are 5\% or less. }
\end{figure}

In H-K vs V-K colour-colour diagrams, NX Pup A and B as well
as NX Pup C lie outside of the region where reddened main sequence stars
with normal extinction would be located. In the J-H vs H-K diagram
NX Pup A falls into the region where Herbig AeBe stars can be found.
NX Pup B and C lie in the region where reddened main sequence stars
and T Tauri stars with low accretion rates and/or face-on disks 
(small IR excess)
are located (Lada and Adams 1992, Strom \ea 1993).
This indicates that both NX Pup A and B have intrinsic IR
excesses which may arise from a circumstellar disk or envelope. 
 Because of the small separation between A and
B, circumstellar material either in form of envelopes or disks should be
strongly disturbed. This has already been pointed out by Henning
et al. (1994) and could explain the non-detection of NX Pup in their
1.3mm continuum search.
The spectral energy distribution
(SED) is plotted in Fig. \ref{sed}. The major part of the system's IR excess
emission --
already noted by Brand \ea (1983) and Reipurth (1983) and fitted by them with a
blackbody of T$_{eff}$ around 1300K -- comes from NX Pup A, as evident in our
Fig. \ref{sed}.

As the reddening vector is almost parallel to the sequence of dwarfs
and giants with normal colours in the H-K vs V-K diagram, we can constrain
the spectral type of NX Pup B. 
If extinction along the line of sight is negligible, we
derive a spectral type earlier than K5 for NX Pup B
and earlier than G6 if A$_V$=1\fm5 as for NX Pup A (Hillenbrand et al.
1992).
The spectrum obtained with the NTT shows the \Ha\ line of NX\,Pup\,AB in
emission with a central absorption dip. 
Other lines in emission are the HeI line
at 587.5nm, the NaI lines at 589.0 and 589.6nm, OI at 630.0nm and
\Hb. Because of the strong saturation of the CCD in the central part
of the spectrum obtained at the NTT, 
we did not try to identify photospheric absorption features. 
The spectrum obtained at the ESO 1.5m shows no evidence for Li I.
Mart\'{\i}n (1994) has estimated that, e.g., in the
combined spectrum of an Herbig AeBe star of spectral type
A2 (50L$_\odot$) and an 
1.5M$_\odot$ T Tauri type star (10L$_\odot$, 10$^6$ yrs) the Li I doublet
should be visible with an equivalent width of 0.02nm.

\begin{table}
\caption{\label{spec}Equivalent width (in nm) of emission ($<0$) and absorption
($>0$) lines in
the spectra of NX\,Pup\,AB and C (28.1.1994, ESO\,1.5m/B\&C; 20.3.1994, NTT/EMMI
 )}
\begin{center}
\begin{minipage}{70mm}
\begin{tabular}{ccc}
\hline
line& NX\,Pup~AB & NX\,Pup~C \\
\hline
\Ha       & -- \footnote{double peaked}  & -2.85 \\
OI 630.0& -0.068&  -0.11  \\
\Hb       & 0.25\footnote{emission core}&  -0.58\\
\Hg       & 0.53   &  -0.17\\ \hline
Li\,I 670.8& -- &  0.054  \\
\end{tabular}
\end{minipage}
\end{center}
\end{table}

The spectral type of NX Pup A has been under debate. Observers in the
UV found systematic earlier spectral types (A0--A2, de Boer 1977, 
Tjin A Djie et al. 1984) than observers in the
visual (F0--F2, Brand et al. 1983, Reipurth 1983). Recently, Blondel
and Tjin A Djie (1994) suggested that the low resolution IUE spectrum
of NX Pup A/B can be decomposed into the spectrum
of a hot boundary layer -- situated between accretion disk and star -- and 
the photospheric spectrum of an F2 type star.

\begin{figure*}[ht]
\centerline{\psfig{figure=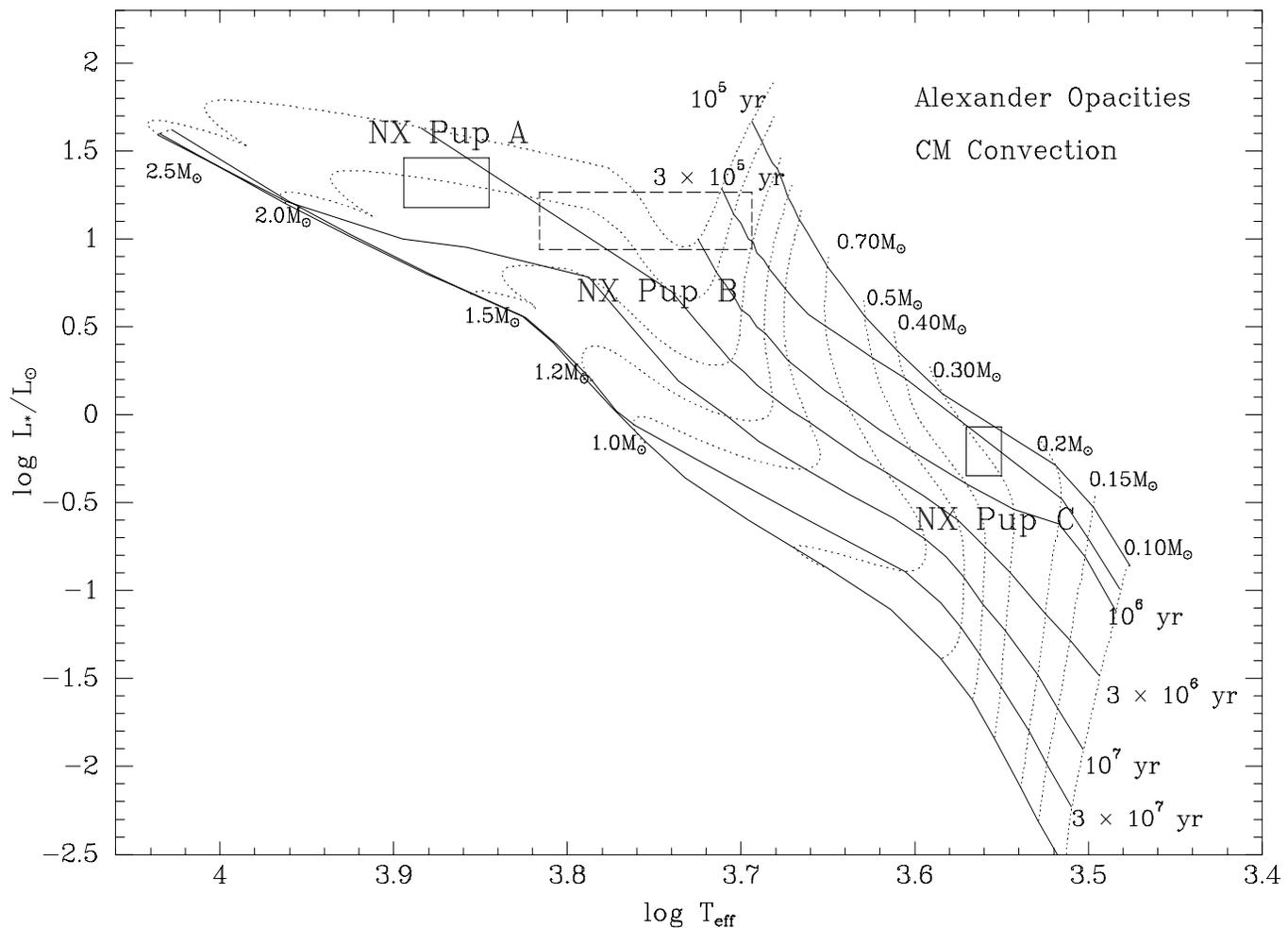,width=19cm,height=14cm}}
\caption{\label{tracks}HR-diagram.
The pre-main sequence evolutionary
tracks are from D'Antona and Mazzitelli (1994).
The solid lines
mark the isochrones and the zero-age main sequence, the dotted lines
are the evolutionary tracks for stars in the mass range from 0.1 to
2.5M$_\odot$.
The positions of NX Pup A, B, and C
in this diagram are marked by boxes.
For NX Pup A we assume a spectral type A7--F2, V=10\fm1--10\fm5, and A$_V$=
0\fm4--0\fm7 (Blondel and Tjin A Djie 1994). For NX Pup B we assume
the same extinction, a spectral type F5--G8, and V=10\fm7--11\fm1.
The evolutionary status of
NX Pup C is better defined, yielding an age around 5$\times$10$^5$yrs
and a mass of 0.30M$_\odot$.}
\end{figure*}

We use the spectral type -- effective temperature calibration
from de Jager and Nieuwenhuijzen (1987) to determine T$_{eff}$.
If we take the extinction values from Blondel and Tjin A Djie (A$_V$ =
0\fm4--0\fm7), assume a spectral type A7--F2 and V=10\fm1--10\fm5
for NX Pup A, we get a luminosity of 15--29 L$_\odot$. 
With the knowledge of the luminosity and the effective temperature
we are able to place NX\,Pup\,A on an HR diagram and to determine its
evolutionary status. New sets of pre-main sequence evolution tracks
have been computed by D'Antona and Mazzitelli
(1994).
From their tracks based on opacities by Alexander et al.
(1989) and the convection model from Canuto and Mazzitelli (1990)
we get a mass of 2M$_\odot$ and an age around $5 \times 10^6$ yr
(see Table \ref{evol} and Fig.\ \ref{tracks}).

The evolutionary status of NX Pup B is less well constrained.
From its position in the colour--colour--diagrams and the fact, that it is
fainter than NX Pup A, a spectral type between mid F and late G
seems to be likely. Assuming V=10\fm7--11\fm1 and the same extinction
values as for NX Pup A, we get a luminosity between 9 and 18 L$_\odot$.
The dashed box in Fig. \ref{tracks} shows that an age from 0.3 to 
$5 \times 10^6$ yrs and a mass between 1.2 and 2.5M$_\odot$ 
is possible. 

Spatially resolved observations of NX Pup A and B redward of
2.2$\mu$m would be desirable. However, because of the 
diffraction limit they have to wait for the next generation of
(southern) telescopes
with their 6 to 8 m mirrors. In the meantime speckle observations
in the R and I band and speckle spectroscopy could provide us
with better estimates on luminosity and spectral type of NX Pup A
and B.

\subsection{NX\,Pup\,C}
 
NX Pup C itself is unresolved on the shift-and-add images. In the
following we regard it as a single star.
 
In the spectra of NX\,Pup\,C the Li\,I 670.8nm absorption line is present
with an equivalent width of 0.054nm (Fig. \ref{li}).
Since Li is depleted in the deeper zones of the convection
layer of low-mass stars, the presence of this line is a sign of youth
and gives a strong hint that NX\,Pup~C is still in its PMS evolution
phase.
 
The equivalent width of the \Ha\ emission is 2.85nm (Table \ref{spec}), 
thus NX\,Pup\,C belongs to the group of classical T\,Tauri stars (CTTS).
TiO bands at 665.2 and 668.1 nm (typical for M-type stars) are visible, 
whereas TiO bands at 559.8, 562.9 and 556.1 nm (present from M2 on)
are not visible.
The spectral A index, which estimates the intensity of the CaH band around
697.5nm (Kirkpatrick \ea 1991), yields a possible MK type
between M0.5V and M4III. 
Taking uncertainties in the extinction correction (see below) into account 
as well as the fact that NX\,Pup~C as a PMS star is still in its contraction 
phase, we get a spectral type between M0.5 and M1.5. 

In order to derive NX Pup C's luminosity, an estimate of the extinction on
the line of sight has to be derived.  This is not straightforward, since we
have three constituents to the extinction: foreground extinction,
extinction within CG1 (radio maps from Harju \ea show that the molecular
cloud extends well beyond NX Pup) and possibly circumstellar extinction.
Another difficulty when estimating the star's luminosity arises from the
presence of both an UV excess (see Table \ref{phot}) and a near-IR excess
(as revealed by the location of NX Pup C in the V-K v. H-K colour-colour 
diagram,
see above). The strong UV excess may be explained by the presence of the
Balmer jump, a feature commonly observed among CTTS (e.g.\ Valenti \ea
1993), while the near-IR excess implies that part of the stellar radiation
is reprocessed within the disk. These excesses and the presence of strong
Balmer lines indicates ongoing accretion in the disk of NX Pup C.
Therefore, the observed total luminosity is the sum of the photospheric
luminosity and the accretion luminosity.

As the maximum contribution of the stellar photosphere to the
total spectral energy distribution (SED) is at about 1$\mu$m (Bertout \ea
1988, Hartigan \ea 1992), the I or J magnitudes are the best measures
of the stellar photospheric flux. The presence of photospheric
absorption lines tells us that the veiling is not too strong. However, 
high-resolution spectra would be necessary for an exact evaluation
of the veiling.

We use the R--I colour and a spectral 
type of M1V for NX\,Pup\,C to compute the extinction.
Colours and bolometric corrections for M dwarfs are from Hartigan
et al. (1994). As interstellar extinction relations are colour-dependent
we use the new values for differential and total extinction
from Grebel and Roberts (1995).
Assuming a normal interstellar extinction law towards NX Pup
and
a distance of 450pc, we derive a total extinction A$_I$=0.93 (A$_V$=1.47) and a
luminosity of 0.85L$_\odot$ for NX\,Pup\,C. If we use the
V--R colour instead of R--I, we get A$_I$=0.43 (A$_V$=0.68) and a
luminosity of 0.45L$_\odot$ for NX\,Pup\,C. This illustrates
the difficulty in deriving
extinction values for stars with abnormal spectral energy
distributions like classical T Tauri stars.

The spectral
energy distribution of NX Pup C is shown in Fig. \ref{sed} and compared 
to that
of an M1 dwarf. The UV and near-IR excess above the photospheric SED are
clearly seen, the latter amounting to $\Delta(K) \approx 0.5$dex, typical 
for classical T Tauri stars (Strom et al. 1989).

We use again the relation
from de Jager and Nieuwenhuijzen (1987) to determine
T$_{eff}$ for NX\,Pup\,C. In the region between M0.5 and M1.5,
T$_{eff}$ is almost independent of the luminosity class. Taking
the uncertainties in the spectral classification into account,
we get an effective temperature between 3550 and 3720K.

From the tracks of D'Antona and Mazzitelli
we get a mass of 0.30M$_\odot$ and an age of $5 \times 10^5$ yrs for NX\,Pup\,C 
(see Table \ref{evol} and Fig.\ \ref{tracks}).

\begin{table}
\caption{\label{evol}Evolutionary status of NX\,Pup\,A, B, and C}
\begin{center}
\begin{minipage}{70mm}
\begin{tabular}{ccccc}
\hline
NX\,Pup    & A    & B & C \\
\hline
sep.      &-- &$0\farcs 128\pm 0\farcs 008$&$6\farcs 98\pm 0\farcs 04$ \\
PA        &-- & $62\fdg 4\pm 5\fdg 7$ &$45\fdg 3\pm 0\fdg 2$ \\
SpT       &A7--F2\footnote{Brand et al. 1983, Reipurth 1983, Blondel \& Tjin A Djie 1994} & F5--G8& M0.5--M1.5\\
L/L$_\odot$ &15--29 & 9--18 & 0.45--0.85 \\
age       &$5 \times 10^6$ yrs & 0.3--$5 \times 10^6$ yrs & $5 \times 10^5$ yrs\\ 
\end{tabular}
\end{minipage}
\end{center}
\end{table}

Our results are sensitive for the choice of evolutionary tracks. Tracks
based on convection models with mixing-length theory yield a 
somewhat higher age for NX Pup C ($10^6$ yrs).

\section{Summary}

We investigated the properties of stars located close to the Herbig Ae/Be
star NX
Pup A. We found that its close companion, NX Pup B, is
very likely a pre-main sequence
star exhibiting strong IR excess, though the lack of a spectrum does not
allow us to assess its mass and age. Nevertheless, its somewhat
lower luminosity than NX Pup A, and its position in colour-colour-diagrams
suggest a spectral type between mid F and late G.
The more distant companion, NX Pup C,
which may or may not be physically associated to NX Pup A/B, is 
identified as a low-mass classical T Tauri star whose age around
$5 \times 10^5$ years
is somewhat younger than the ages deduced by others for NX\,Pup~AB and CG1.
This provides the first evidence that low-mass star formation occurred in
this cometary globule.

\acknowledgements
We thank P.S. Th\'e and the LTPV program for the permission
to use their photometric data on NX Pup prior to publication
and A. Tokovinine for providing IRAF scripts he
developed to perform photometry of close binaries.
We would like to thank Thomas Lehmann for providing his measurements of the
shutter delay time of the CCD-camera at the Danish 1.5m telescope.
Discussions with R. Wilson are gratefully acknowledged.
 
WB and EKG were supported by student fellowships of the European Southern
Observatory. 

This research has made use of the Simbad database,
operated at CDS, Strasbourg, France, NASA's Astrophysics Data System (ADS),
version 4.0, and the IRAF PACKAGE C128, developed by E.\ Tessier at the
Observatory of Grenoble.

\enddocument